\documentclass[prd,showpacs]{revtex4}
\begin{document}

\title{Symmetry breaking in non conservative systems}

\author{N.E. Mart\'{\i}nez-P\'erez}
 \email{nephtalieliceo@hotmail.com}
  \affiliation{Benem\'erita Universidad Aut\'onoma de Puebla, Facultad de Ciencias
F\'{\i}sico Matem\'aticas, P.O. Box 165, 72570 Puebla, M\'exico.}
 
\author{C. Ram\'{\i}rez}
 \email{cramirez@fcfm.buap.mx}
 \affiliation{Benem\'erita Universidad Aut\'onoma de Puebla, Facultad de Ciencias
F\'{\i}sico Matem\'aticas, P.O. Box 165, 72570 Puebla, M\'exico.}

 \date{\today}

\begin{abstract}
We apply Noether's theorem to show how the invariances of conservative systems are broken for nonconservative systems, in the variational formulation of Galley. This formulation considers a conservative action, extended by the inclusion of a time reversed sector and a nonconservative generalized potential. We assume that this potential is invariant under the symmetries of the initial conservative system. The breaking occurs because the time reversed sector requires inverse symmetry transformations, under which the nonconservative potential is not invariant. The resulting violation of the conservation laws is consistent with the equations of motion. We generalize this formulation for fermionic and sypersymmetric systems. In the case of a supersymmetric oscillator, the effect of damping is that the bosonic and fermionic components become different frequencies. Considering that initially the nonconservative action is invariant under supersymmetry, and that the breaking is associated to an instability, this result is reminiscent of spontaneous symmetry breaking.
\end{abstract}
\pacs{04.20.Fy,12.60.Jv}
\maketitle

\section{Introduction}
The study of a general mechanical system includes influences of external factors, whose origin and detailed description may be partially or fully unknown. The evolution of such systems is frequently irreversible and non invariant under time reversal. There are fundamental questions related to this issue, like time direction, the second principle of thermodynamics, etc. A formulation for such systems has been given in terms of equations of motion since a long time. 

Hamilton's principle gives a way to obtain the equations of motion of conservative systems by the variation of the action, with the variables fixed at the initial and final times. The physical trajectory goes along a curve through these initial and final points and minimizes the action. Thus, this trajectory is determinated by conditions in the past and in the future, which is natural for time reversible systems. The symmetries of these systems lead to conservation laws, which can be obtained from Noether's theorem. 

The application of nonconservative forces leads to the violation of energy conservation, and could violate also other symmetries. 
If one wishes to describe nonconservative systems, not invariant under time reversal, independent boundary conditions in the past and in the future might be contradictory. A solution to this problem has been proposed long time ago by Schwinger, by the inclusion of a different time reversed dynamics \cite{schwinger}, which corresponds to the doubling of variables of older works \cite{bateman}. 
In a recent proposal by Galley \cite{galley1,galley2}, a systematic approach has been given for a lagrangian formulation for nonconservative systems, which amounts to a doubling of the variables, by the inclusion of a time reversed sector and a nonconservative generalized potential, along with a modification of Hamilton's variational principle. This proposal is similar to the closed time path formulation, see e.g. \cite{chou}. A similar development for classical and quantum mechanics was given previously by Polonyi in \cite{polonyi}, who in a recent paper consider the issue of breaking of time reversal symmetry \cite{polonyi2}. In the proposal of Galley, there are boundary conditions only at the initial time, and at the final time the corresponding variables coincide. Once the equations of motion are derived, a so called physical limit is applied, which amounts to the identification of both sets of variables, from which the nonconservative equations of motion follow.

Classical mechanics deals with the evolution of systems described by quantities like the position, angles, and so on. However, if we see it as a limit of quantum mechanics, it is natural to ask about fermionic degrees of freedom, which describe the matter in the universe, and whose properties are properly described by anticommuting quantities, in order to attain Fermi-Dirac statistics. This has lead to the formulation of classical fermionic mechanics \cite{casalbuoni}, whose quantization can be done in the canonical formulation, and strictly requires the Hamiltonian Dirac formulation \cite{teitelboim}.  In fact, as classical fermionic variables do not correspond to classical observables, they are mathematical constructions which have proven to be useful in the formulation of fermionic, hence quantum, theories. For instance, from the relativistic classical action of a fermionic point particle, the Dirac equation can be obtained \cite{teitelboim}. Otherwise, fermionic field theory is formulated starting from classical actions, which are also required for the path integral formulation.

Supersymmetry has been developed as a symmetry towards the unification of the fundamental interactions. It transforms bosons into fermions and its basic entities are supermultiplets, which contain bosons and fermions, see e.g. \cite{wess}. On the other side, supersymmetry imposes strong restrictions, like the cancellation of divergences in field theory, one of the main reasons of its success, or the equality of the masses of particles in the same supermultiplet, one of the principal objections against it, which has lead to the requirement that supersymmetry is broken, usually spontaneously. 
Supersymmetric mechanics \cite{bellucci} can be realized by an extension of time translations, by means of transformations with anticommutative parameters. Thus, supersymmetry is intimately related to energy conservation, and a question which arises naturally is what happens with nonconservative systems, which should broken it. On the other side, if a supersymmetric system is exposed to external forces, these forces could be due to supersymmetric unknown factors. In this case, this theory should have a supersymmetric structure. 

In this paper, we formulate Noether's theorem for the nonconservative action of Galley \cite{galley1}. This action can be straightforwardly constructed to have the same invariances as the original conservative action. However, considering the time reversed sector, we make the ansatz that the symmetry transformations act on this sector by their inverses. This has the consequence that the nonconservative potential breaks these symmetries. In this way, from Noether's theorem follows the violation of the conservation laws corresponding to all the symmetries of the original conservative action.   Further, we consider fermionic systems, whose boundary conditions are determined by only one parameter; hence their nonconservative generalization requires a slight modification of the boundary conditions. As a last step, we give a straightforward generalization for supersymmetric systems, which is done in the superspace formalism. Thus, the nonconservative potential is a superfield; hence it is supersymmetric by construction. Along the paper we consider the example of the damped oscillator. It is remarkable that in the supersymmetric case, where the bosonic and the fermionic variables have the same frequency in the conservative theory, now these variables have different frequencies, as a consequence of the nonconservative interactions. This situation is reminiscent the situation of spontaneous supersymmetry breaking in field theory, where the initial action is supersymmetric as well, and after a breaking due to instability, the field components become different masses  \cite{wess}.
In Section II we give a short review of the formulation of Galley and show Noether's theorem for the nonconservative action. In section III we consider the generalization for fermionic systems. In Section IV we perform the supersymmetric generalization. In the last Section we draw some conclusions.

\section{Lagrangian Formulation}

Hamilton's principle in mechanics establishes that the trajectory of a system in an arbitrary time interval $[t_{i},t_{f}]$, on which act conservative forces, is an extremum of the action. The variation is done on curves which go through two fixed points, at the initial and final times. Instead of it, we could consider curves that go at the initial time through a fixed point, with a fixed velocity. In this case the point at the final time should be fixed also, although it cannot be arbitrary, as it is determined by the initial position and velocity. It can be seen that in this way we obtain the Euler-Lagrange equations as well \cite{arnold}. As far as the kinetic and potential terms do not depend explicitly on time, the E-L equations are symmetric under time reversal. A solution can violate this symmetry due to the choice of initial conditions, however its time inverted is also a solution, i.e. the space of solutions is symmetric. 

The lagrangian formulation of Galley supposes that there is a conservative lagrangian $L(q,\dot{q})$, where $q$ are in general $n$-dimensional vectors. Nonconservativity is attained by a doubling of the degrees of freedom $q\rightarrow(q_1,q_2)$, and the variational principle is modified in such a way that $q_{2}(t)$ effectively runs back in time, and appears as a continuation of $q_{1}(t)$. Thus the boundary conditions for the variation are that, at the initial time both variables are fixed, and at the final time their values coincide, but have an arbitrary variation. Once the variation done, the doubling is reverted, at the level of the equations of motion, by the so called physical limit $q_1=q_2=q$. This is similar to the Closed Path-Time approach \cite{schwinger,chou}. In fact, both variables could be arranged as one, beginning and finishing at $t_i$, after a closed time path $t_i\rightarrow t_f\rightarrow t_i$ \cite{polonyi}. 
Galley avoids the time loop and instead of it modifies the setting of the action
\begin{equation}
S=\int_{t_i}^{t_f}L({q}_1,{\dot{q}}_1)dt+\int_{t_f}^{t_i}L({q}_2,{\dot{q}}_2)dt=\int_{t_i}^{t_f}L({q}_1,{\dot{q}}_1)dt-\int_{t_i}^{t_f}L({q}_2,{\dot{q}}_2)dt.
\end{equation}
This setting allows to add to the action a nonconservative ``generalized potential", $K(q_1,\dot q_1,q_2,\dot q_2)$, which depends on both variables and is antisymmetric under $1\leftrightarrow 2$ 
\begin{equation}
K(q_1,\dot q_1,q_2,\dot q_2)=-K(q_2,\dot q_2,q_1,\dot q_1).\label{antisym}
\end{equation}
Thus, te nonconservative action is given by
\begin{equation}
S=\int_{t_i}^{t_f}\Lambda(q_1,\dot q_1,q_2,\dot q_2)dt=\int_{t_i}^{t_f}\left[L({q}_1,{\dot{q}}_1)-L({q}_2,{\dot{q}}_2)+K(q_1,\dot q_1,q_2,\dot q_2)\right]dt.\label{ncl}
\end{equation}
The variation is given now with the boundary conditions that at the initial time $\delta{q}_1(t_i)=\delta{q}_2(t_i)=0$, and at the final time both variables coincide ${q}_1(t_f)={q}_2(t_f)$, and have arbitrary variations $\delta{q}_1(t_f)=\delta{q}_2(t_f)$. Aditionally, ${\dot {q}}_1(t_f)={\dot {q}}_2(t_f)$. Thus
\begin{eqnarray}
\delta S&=&\int_{t_{i}}^{t_{f}}\delta\Lambda(q_1,\dot q_1,q_2,\dot q_2)dt
=\left[\delta q_{1}\left(\frac{\partial L}{\partial{\dot q}_{1}}+\frac{\partial K}{\partial{\dot q}_{1}}\right)+\delta q_{2}\left(-\frac{\partial L}{\partial{\dot q}_{2}}+\frac{\partial K}{\partial{\dot q}_{2}}\right)\right]_{t=t_{f}}\nonumber\\
&+&\int_{t_{i}}^{t_{f}}\left\{\delta q_{1}\left[\frac{\partial}{\partial q_{1}}(L+K)-\frac{d}{dt}\frac{\partial}{\partial{\dot q}_{1}}(L+K)\right]-\delta q_{2}\left[\frac{\partial}{\partial q_{2}}(L-K)-\frac{d}{dt}\frac{\partial}{\partial{\dot q}_{2}}(L-K)\right]\right\}dt.
\end{eqnarray}
The boundary terms vanish after taking into account the boundary conditions and the antisymmetry of $K$, from which in particular follows
\begin{equation}
\left.\left(\frac{\partial K}{\partial{\dot q}_{1}}+\frac{\partial K}{\partial{\dot q}_{2}}\right)\right|_{q_1=q_2,\ \dot q_1=\dot q_2}=0.\label{derk}
\end{equation}
Thus, the equations of motion are
\begin{eqnarray}
\left(\frac{\partial}{\partial q_{1}}-\frac{d}{dt}\frac{\partial}{\partial{\dot q}_{1}}\right)\left[L(q_1,\dot{q_1})+K(q_1,\dot{q_1},q_2,\dot{q_2})\right]=0,\label{eq1}\\
\left(\frac{\partial}{\partial q_{2}}-\frac{d}{dt}\frac{\partial}{\partial{\dot q}_{2}}\right)\left[L(q_2,\dot{q_2})-K(q_1,\dot{q_1},q_2,\dot{q_2})\right]=0.\label{eq2}
\end{eqnarray}
Due to (\ref{derk}), in the physical limit  $q_{1}=q_{2}=q$, the second equation is identical with the first one. Thus, the final nonconservative equations of motion are
\begin{eqnarray}
\frac{\partial L}{\partial q}-\frac{d}{dt}\frac{\partial L}{\partial{\dot q}}
=-F_{K}\equiv\left.\left(\frac{\partial}{\partial q_{2}}-\frac{d}{dt}\frac{\partial}{\partial{\dot q}_{2}}\right)K(q_{1},\dot{q}_{1},q_{2},\dot{q}_{2})\right\vert_{q_{1}=q_{2}=q},\label{ncfuerzas}
\end{eqnarray}
where $F_K$ are the nonconservative forces \cite{galley2}. Therefore, if $K=0$, (\ref{ncfuerzas}) are the usual equations of motion. This formalism is straightforwardly generalized for any number of degrees of freedom, for any conservative Lagrangian \cite{galley1}.

\subsection{Symmetry breaking}
In \cite{galley2}, from Noether's theorem for the conservative system, the violation of the corresponding conservation laws is derived, considering the nonconservative equations of motion (\ref{eq1}) and (\ref{eq2}). In fact, it is easy to see that a direct application of Noether's theorem to the nonconservative action (\ref{ncl}) gives identically vanishing results. We 
assume that the nonconservative potential $K(q_1,\dot q_1,q_2,\dot q_2)$ has the same invariances as the conservative Lagrangian, i.e. if $t\rightarrow t'=t+\delta t$, and $q\rightarrow q'=q+\delta t\dot q+\delta_\alpha q$, where $\delta_\alpha q$ corresponds to "internal" symmetries; then $\delta_t K(q_1,\dot q_1,q_2,\dot q_2)=\delta t\frac{d}{dt}K(q_1,\dot q_1,q_2,\dot q_2)$, and $\delta_\alpha K(q_1,\dot q_1,q_2,\dot q_2)=0$. However, considering the characteristics of the two sectors of the nonconservative action (\ref{ncl}), we make the ansatz that they transform as
\begin{eqnarray}
{\rm a) \ For }\ q_1:\ t\rightarrow t'=t+\delta t\ \ {\rm and }\ \ q_1\rightarrow q_1'=q_1+\delta t\dot q_1+\delta_\alpha q_1.\label{itema}\\
{\rm b) \ For }\ q_2:\ t\rightarrow t'=t-\delta t\ \ {\rm and }\ \ q_2\rightarrow q_2'=q_2-\delta t\dot q_2-\delta_\alpha q_2.\label{itemb}
\end{eqnarray}
The two conservative actions in (\ref{ncl}) are invariant under these transformations, but not the nonconservative potential, which transforms as
\begin{eqnarray}
\delta K
=\delta t\left[\frac{dK}{dt}-2\left(\dot q_{2}\frac{\partial K}{\partial q_2}+\ddot q_{2}\frac{\partial K}{\partial \dot q_2}\right)\right]-2\left(\delta_\alpha q_{2}\frac{\partial K}{\partial q_2}+\delta_\alpha\dot q_{2}\frac{\partial K}{\partial \dot q_2}\right).\label{deltak}
\end{eqnarray}
Noether's theorem is usually formulated from a variation of the action. However, considering that in this variation the boundary conditions play no role, it can be formulated directly from the Lagrangian. Indeed, for $L(q,\dot q)$ invariant under the transformations (\ref{itema}), we have
$\delta L=\delta t\frac{dL}{dt}$, which can be written as
\begin{equation}
\delta L=\delta t\left(\dot q\frac{\partial L}{\partial q}+\ddot q\frac{\partial L}{\partial\dot q}\right)+\delta_\alpha q\frac{\partial L}{\partial q}+\delta_\alpha\dot q\frac{\partial L}{\partial\dot q}=\frac{d}{dt}\left[(\delta t\dot q+\delta_\alpha q)\frac{\partial L}{\partial\dot q}\right]+(\delta t\dot q+\delta_\alpha q)\left(\frac{\partial L}{\partial q}-\frac{d}{dt}\frac{\partial L}{\partial\dot q}\right).
\end{equation}
Hence, equating the right hand side of both equations, and taking into account the equations of motion, we get
\begin{equation}
\frac{d}{dt}\left[\delta t\dot q\left(\frac{\partial L}{\partial\dot q}-L\right)+\delta_\alpha q\frac{\partial L}{\partial\dot q}\right]=0.
\end{equation}
In the same way, the variation of (\ref{ncl}) gives
\begin{equation}
\delta \Lambda=\delta t\left(\frac{dL_1}{dt}+\frac{dL_2}{dt}\right)+\delta K,\label{dlambda1}
\end{equation}
where $\delta K$ is given by (\ref{deltak}), $L_1\equiv L(q_1,\dot q_1)$, and $L_2\equiv L(q_2,\dot q_2)$. Further, from (\ref{dlambda1}), modulo the equations of motion (\ref{eq1}) and (\ref{eq2}), we get
\begin{equation}
\delta \Lambda=\frac{d}{dt}\left[\delta t\left(\dot q_1\frac{\partial\Lambda}{\partial\dot q_1}-\dot q_2\frac{\partial\Lambda}{\partial\dot q_2}\right)+\delta_\alpha q_1\frac{\partial\Lambda}{\partial\dot q_1}-\delta_\alpha q_2\frac{\partial\Lambda}{\partial\dot q_2}\right].\label{dlambda}
\end{equation}
Therefore (\ref{dlambda}) and (\ref{dlambda1}) give, after some rearrangements and using (\ref{deltak})
\begin{eqnarray}
&&\frac{d}{dt}\left\{\delta t\left[\dot q_1\frac{\partial (L_1+K)}{\partial\dot q_1}+\dot q_2\frac{\partial(L_2+K)}{\partial\dot q_2}-L_1-L_2-K\right]+\delta_\alpha q_1\frac{\partial (L_1+K)}{\partial\dot q_1}+\delta_\alpha q_2\frac{\partial (L_2+K)}{\partial\dot q_2}\right\}\nonumber\\
&&=-2\left(\delta t\dot q_{2}+\delta_\alpha q_{2}\right)\left(\frac{\partial K}{\partial q_2}-\frac{d}{dt}\frac{\partial K}{\partial \dot q_2}\right),\label{noncec}
\end{eqnarray}
from which in the physical limit we obtain
\begin{eqnarray}
\frac{dE}{dt}&=&\frac{d}{dt}\left(\dot q\frac{\partial L}{\partial\dot q}-L\right)=\dot qF_K,\label{nce1}\\
\frac{dJ_\alpha}{dt}&=&\frac{d}{dt}\left(\delta_\alpha q\frac{\partial L}{\partial\dot q}\right)=\delta_\alpha qF_K,\label{nce2}
\end{eqnarray}
consistently with the equations of motion (\ref{ncfuerzas}), and which coincide with the results of \cite{galley1}. 

Therefore, for the nonoconservative theories that can be represented by Galley's approach, with a scalar nonconservative potential $K$, the invariances of the original conservative action are broken. Moreover, the implementation of Noether's theorem for the Lagrangian $\Lambda$ in this framework, leads consistently to the equations (\ref{nce1}) and (\ref{nce2}). Note that in both cases, it is the same nonconservative force which leads to the violation. All the results of this section can be straightforwardly generalized for any number of degrees of freedom.

\subsection{Damped oscillator}
As an example we consider the damped oscillator \cite{galley1}. The conservative lagrangian is the one of the oscillator $L(q,\dot q)=m/2(\dot q^{2}-\omega^{2}q^{2})$ and $K(q_{1},\dot q_{1},q_{2},\dot q_{2})=c/2(q_{1}\dot q_{2}-q_{2}\dot q_{1})$. Hence
\begin{eqnarray}
\Lambda(q_1,q_2,\dot q_1,\dot q_2)=\frac{m}{2}(\dot q_1^{2}-\omega_0^{2}q_1^{2})-\frac{m}{2}(\dot q_2^{2}-\omega_0^{2}q_2^{2})-\frac{c}{2}(q_{1}\dot q_{2}-q_{2}\dot q_{1}).
\end{eqnarray}
The equation of motion, in the physical limit is
\begin{equation}
m\ddot q+c\dot q+m\omega_0^2q=0.\label{eqq}
\end{equation}
In fact, this equation has been obtained from a variational principle already long ago \cite{bateman}, where a doubling of variables has been already required. Equation (\ref{noncec}) is in this case
\begin{equation}
\frac{1}{2}\frac{d}{dt}\left[\frac{m}{2}({\dot q}_1^2+\omega_0^2 q_1^2)+\frac{m}{2}({\dot q}_2^2+\omega_0^2 q_2^2)\right]=-c\dot q_1\dot q_2, \label{enero}
\end{equation}
i.e. in the physical limit 
\begin{equation}
\frac{dE}{dt}=-\frac{2c}{m}E_{\rm cin}.\label{enerop}
\end{equation}
Let us now consider a damped 3-dimensional oscillator
\begin{eqnarray}
\Lambda(\vec q_1,\vec q_2,\dot{\vec q}_1,\dot{\vec q}_2)=\frac{m}{2}({\dot{\vec q}_1}^{2}-\omega_0^{2}{\vec{q}_1}^{2})-\frac{m}{2}(\dot{\vec q}_2^{2}-\omega_0^{2}\vec q_2^{2})-\frac{c}{2}(\vec q_{1}\dot{\vec q}_{2}-\vec q_{2}\dot{\vec q}_{1}). 
\end{eqnarray}
In this case $\delta_\alpha q_i=\epsilon_{ijk}\alpha_iq_k$, and from (\ref{noncec}), additionally to (\ref{enero}) we get
\begin{equation}
\frac{1}{2}\frac{d\vec J}{dt}=\frac{1}{2}\frac{d}{dt}\left[m\left(\vec q_1\times\dot{\vec{q}}_1+\vec q_2\times\dot{\vec{q}}_2\right)-c\vec q_1\times\vec{q}_2\right]=-c\vec q_2\times\dot{\vec{q}}_1,
\end{equation}
which in the physical limit gives the violation of the angular momentum conservation law
\begin{equation}
\frac{d\vec L}{dt}=-\frac{c}{m}\vec L.
\end{equation}
\section{Fermionic systems}
Fermionic systems in classical mechanics are formulated by means of anticommuting variables. We consider variables $\psi$ and $\bar\psi$, where $\bar\psi$ is the complex conjugated. These variables can have an $n$-dimensional index and satisfy $\psi_{i}\psi_{j}=-\psi_{j}\psi_{i}$, $\psi_{i}\bar\psi_{j}=-\bar\psi_{j}\psi_{i}$, and $\bar\psi_{i}\bar\psi_{j}=-\bar\psi_{j}\bar\psi_{i}$, otherwise they commute with bosonic quantities. Thus they are nilpotent, $\psi_{i}\psi_{i}=0$. Complex conjugation reverses the order like hermitian conjugation, thus giving a sign, e.g. $(\psi_{i}\psi_{j})^{\dagger}=-\bar\psi_{i}\bar\psi_{j}$. In the following we will not indicate the $n$-dimensional indices. We will adopt the convention that fermionic derivatives act on the left, as
\begin{equation}
\frac{\partial}{\partial\psi_i} \psi_j=\delta_{ij}-\psi_j\frac{\partial}{\partial\psi_i}.
\end{equation}

The kinetic term of fermionic systems is first order, which means that the trajectory is determined by fixing only one parameter. For this reason, in \cite{teitelboim} a boundary term has been proposed for fermionic actions. Consider for example a system described by one fermionic variable $\psi(t)$ with lagrangian $L(\psi,\bar\psi)=i/2(\bar\psi\dot\psi+\psi\dot{\bar\psi})+\psi\bar\psi$; we use an economical notation, where the dependence of the Lagrangian on the first time derivatives is not indicated. According to \cite{teitelboim}, to the action are added suitable boundary terms
\begin{equation}
S=\int_{t_{i}}^{t_{f}}\left[\frac{i}{2}(\bar\psi\dot\psi+\psi\dot{\bar\psi})-\omega\psi\bar\psi\right]dt-
\frac{i}{2}\left[\psi(t_{1})\bar\psi(t_{2})+\bar\psi(t_{1})\psi(t_{2})\right],
\end{equation}
whose variation with the boundary conditions $\delta\psi(t_{1})+\delta\psi(t_{2})=0$ and $\delta\bar\psi(t_{1})+\delta\bar\psi(t_{2})=0$, gives the expected equations of motion $i\dot\psi+\omega\psi=0$ and $i\dot{\bar\psi}-\omega\bar\psi=0$. However, this procedure depends on the action and seems to be incompatible with supersymmetry, as it affects only the fermionic variables. 

Instead of adding boundary terms, we proceed following the observations at the beginning of the previous section. Thus we fix the values of the fermionic coordinates at the initial and final times, with the consistency restriction that one of both values is determined by the other. As long as Hamilton's principle requires only that the variations of the coordinates at the initial and final times vanish, there is no problem if we do not bother on this dependence, which in fact turns out after solving the equations of motion.
On the other side, a nonconservative formulation for fermions, after the doubling of the variables $\psi\rightarrow(\psi_{1},\psi_{2})$ and $\bar\psi\rightarrow(\bar\psi_{1},\bar\psi_{2})$, requires that $\psi_{1}(t_{f})=\psi_{2}(t_{f})$, $\bar\psi_{1}(t_{f})=\bar\psi_{2}(t_{f})$. This means that for consistency, also the values at the initial time should coincide $\psi_{1}(t_{i})=\psi_{2}(t_{i})$, $\bar\psi_{1}(t_{i})=\bar\psi_{2}(t_{i})$. Furthermore, a general form of the nonconservative potential requires the conditions $\dot\psi_{1}(t_{f})=\dot\psi_{2}(t_{f})$ and $\dot{\bar{\psi}}_{1}(t_{f})=\dot{\bar{\psi}}_{2}(t_{f})$.
Therefore the nonconservative action is \cite{previo}
\begin{equation}
S=\int_{t_{i}}^{t_{f}}\left[L(\psi_{1},\bar\psi_{1})-L(\psi_{2},\bar\psi_{2})+K(\psi_{1},\psi_{2},\bar\psi_{1},\bar\psi_{2})\right]dt,
\end{equation}
with $K(\psi_{1},\psi_{2},\bar\psi_{1},\bar\psi_{2})$ antisymmetric under $1\leftrightarrow 2$. The boundary conditions are $\delta\psi_{a}(t_{i})=\delta\bar\psi_{a}(t_{i})=0$ $(a=1,2)$, $\psi_{1}(t_{i,f})=\psi_{2}(t_{i,f})$, $\bar\psi_{1}(t_{i,f})=\bar\psi_{2}(t_{i,f})$, $\delta\psi_{1}(t_{f})=\delta\psi_{2}(t_{f})$ and $\delta\bar\psi_{1}(t_{f})=\delta\bar\psi_{2}(t_{f})$. Following the same steps as in the preceding section, and considering the relations corresponding to (\ref{derk}),
we get the same equations of motion as in (\ref{eq1}) and (\ref{eq2})
\begin{eqnarray}
\frac{\partial}{\partial \psi_{1}}(L+K)-\frac{d}{dt}\frac{\partial}{\partial{\dot\psi}_{1}}(L+K)=0,\label{eqf1}\\
\frac{\partial}{\partial \psi_{2}}(L-K)-\frac{d}{dt}\frac{\partial}{\partial{\dot\psi}_{2}}(L-K)=0,\label{eqf2}
\end{eqnarray}
and their complex conjugated, with $\psi\rightarrow\bar\psi$. Again, these equations coincide in the physical limit $\psi_{1}=\psi_{2}=\psi$ and $\bar\psi_{1}=\bar\psi_{2}=\bar\psi$. Hence the nonconservative equations of motion are
\begin{eqnarray}
\frac{\partial L}{\partial \psi}-\frac{d}{dt}\frac{\partial L}{\partial{\dot\psi}}
&=&-\Phi_{K}\equiv\left.\left(\frac{\partial}{\partial\psi_{2}}-\frac{d}{dt}\frac{\partial}{\partial{\dot\psi}_{2}}\right)K(\psi_{1},\bar{\psi}_{1},\psi_{2},\bar{\psi}_{2})\right\vert_{\psi_{1}=\psi_{2}=\psi},\\
\frac{\partial L}{\partial \bar{\psi}}-\frac{d}{dt}\frac{\partial L}{\partial{\dot{\bar{\psi}}}}
&=&-\bar \Phi_{K}\equiv\left.\left(\frac{\partial}{\partial\bar{\psi}_{2}}-\frac{d}{dt}\frac{\partial}{\partial{\dot{\bar{\psi}}}_{2}}\right)K(\psi_{1},\bar{\psi}_{1},\psi_{2},\bar{\psi}_{2})\right\vert_{\psi_{1}=\psi_{2}=\psi},
\end{eqnarray} 
where $\Phi_K$ and $\bar\Phi_K$ are the nonconservative fermionic forces.

Regarding Noether's theorem, the variation of the Lagrangian works in the same way as in the bosonic case, considering that for the conservative Lagrangian $L(\psi,\dot\psi)$
\begin{equation}
\delta L=\delta t\left(\dot\psi\frac{\partial L}{\partial\psi}+\dot{\bar\psi}\frac{\partial L}{\partial\bar\psi}+\ddot\psi\frac{\partial L}{\partial\dot\psi}+\ddot{\bar\psi}\frac{\partial L}{\partial\dot{\bar\psi}}\right)+\delta_\alpha\psi\frac{\partial L}{\partial\psi}+\delta_\alpha \bar\psi\frac{\partial L}{\partial\bar\psi}+\delta_\alpha\dot\psi\frac{\partial L}{\partial\dot\psi}+\delta_\alpha\dot{\bar\psi}\frac{\partial L}{\partial\dot{\bar\psi}}=\delta t\frac{dL}{dt}.
\end{equation}
Thus, the equations corresponding to (\ref{itema})-(\ref{nce2}) are the same.
\subsection{Damped fermionic oscillator}
Let us now consider a damped fermionic oscillator. For the lagrangian we consider the fermionic oscillator 
\begin{equation}
L(\psi,\bar\psi)=-i\frac{m}{2}\left(\bar\psi\dot\psi+\psi\dot{\bar\psi}\right)+m\omega_0\psi\bar\psi,
\end{equation}
and 
\begin{equation}
K(\psi_{1},\bar\psi_{1},\psi_{2},\bar\psi_{2})=\frac{\mu}{2}\left(\psi_{1}\dot{\bar\psi}_{2}-\psi_{2}\dot{\bar\psi}_{1}-\bar\psi_1\dot\psi_2+\bar\psi_2\dot\psi_1\right).
\end{equation}
Hence $\Lambda=L(\psi_{1},\bar\psi_{1})-L(\psi_{2},\bar\psi_{2})+K(\psi_{1},\bar\psi_{1},\psi_{2},\bar\psi_{2})$.
The equations of motion are 
$i\dot\psi_1+\omega_0\psi_1+\frac{\mu}{m}\dot\psi_2=0$,
$-i\dot{\bar\psi}_1+\omega_0\bar\psi_1+\frac{\mu}{m}\dot{\bar\psi}_2=0$,
i.e. in the physical limit
\begin{eqnarray}
i\dot{\psi}+\zeta\psi&=&0, \label{eqfnc1}\\
i\dot{\bar{\psi}}-\bar{\zeta}\bar\psi&=&0, \label{eqfnc2}
\end{eqnarray}
where the frequency is complex and is given by
\begin{equation}
\zeta=\frac{\omega_0}{1+\frac{\mu^{2}}{m^2}}\left(1+i\frac{\mu}{m}\right).\label{cfrec}
\end{equation}
The energy change rate is obtained from (\ref{noncec}),
$\frac{m\omega_0}{2}\frac{d}{dt}(\psi_1\bar\psi_1+\psi_2\bar\psi_2)=-\mu(\dot{\psi}_2\dot{\bar{\psi}}_1+\dot{\psi}_1\dot{\bar{\psi}}_2)$, which in the physical limit gives
\begin{equation}
m\omega_0\frac{d}{dt}(\psi\bar\psi)=-2\mu\dot{\psi}\dot{\bar{\psi}},
\end{equation}
consistently with the solutions $\psi(t)=e^{i\zeta t}\psi_{0}$ and $\bar\psi(t)=e^{-i\bar{\zeta}t}\bar\psi_{0}$.

\section{Supersymmetric formulation}
Supersymmetric mechanics can be realized by an extension of time to a Grassmann space, or superspace, $t\rightarrow z\equiv(t,\theta,\bar\theta)$, where $\theta$ and $\bar{\theta}$ are anticommuting variables. There are derivatives defined in these spaces by the rules $\{\partial_\theta,\theta\}=1$, $\{\partial_{\bar\theta},\bar\theta\}=1$, $\{\partial_\theta,\bar\theta\}=0$ and $\{\partial_{\bar\theta},\theta\}=0$, and integration
$\int d\theta=0$, $\int d\theta\theta=1$, $\int d\bar\theta=0$, $\int d\bar\theta\bar\theta=-1$. The dynamical variables are extended to superfields $q(t)\rightarrow \phi(t,\theta,\bar\theta)$, which are real and are given by a finite expansion in the fermionic variables as 
\begin{equation}
\phi(t,\theta,\bar{\theta})=q(t)+\theta\psi(t)-\bar{\theta}\bar{\psi}(t)+\theta\bar{\theta}b(t),\label{superf}
\end{equation}
where the variables $q(t), \psi(t), \bar\psi(t)$ and $b(t)$ are called the components of the superfield. In the following, for a generic superfield $\phi$, we will also denote $q=\phi_q$, $\psi=\phi_\theta$, $\bar\psi=\phi_{\bar\theta}$ and $b=\phi_{\theta\bar\theta}$. Note that
$\theta\bar{\theta}$ is real. The transformations of supersymmetry are generated by the fermionic charges $Q=\frac{d}{d\theta}-i\bar{\theta}\frac{d}{dt}$
and $\bar{Q}=-\frac{d}{d\bar{\theta}}+i\theta\frac{d}{dt}$, which satisfy
$\{Q,\bar{Q}\}\equiv Q\bar Q+\bar QQ=2i\frac{d}{dt}$. Thus a supersymmetric transformation is written as $\delta_\xi\phi(t,\theta,\bar\theta)=(\xi Q-\bar\xi\bar Q)\phi(t,\theta,\bar\theta)=\xi\psi-\bar\xi\bar\psi+\theta\bar\xi(b+i\dot q)-\bar\theta\xi(b-i\dot q)+i\theta\bar\theta(\xi\dot\psi+\bar{\xi}\dot{\bar{\psi}})$, from which, by comparison of the components, turn out the infinitesimal transformations
\begin{equation}
\delta_\xi q=\xi\psi-\bar\xi\bar\psi, \quad\delta_\xi\psi=\bar\xi(b+i\dot q), \quad\delta_\xi\bar\psi=\xi(b-i\dot q) \quad{\rm and}\quad \delta_\xi b=i(\xi\dot\psi+\bar{\xi}\dot{\bar{\psi}}),\label{susyt}
\end{equation}
or in finite form
\begin{equation}
\phi'(t,\theta,\bar\theta)=\phi(t-i\xi\bar\theta-i\bar\xi\theta,\theta+\xi,\bar\theta+\bar\xi).
\end{equation}
Thus supersymmetry transformations can be seen as a certain type of superspace translation.
There are covariant derivatives $D=\frac{d}{d\theta}+i\bar{\theta}\frac{d}{dt}$ and $\bar{D}=-\frac{d}{d\bar{\theta}}-i\theta\frac{d}{dt}$, which satisfy $\{D,\bar{D}\}=-2i\frac{d}{dt}$, $\{Q,D\}=\{Q,\bar{D}\}=\{\bar Q,D\}=\{\bar Q,\bar D\}=0$. Thus, quantities obtained from superfields by the action of covariant derivatives will transform as superfields. The integral of the covariant derivatives of a superfield 
\begin{eqnarray}
D\phi(t,\theta,\bar\theta)&=&\psi+\bar\theta(i\dot{q}+b)-i\theta\bar{\theta}\dot{\psi},\label{covd}\\
\bar D\phi(t,\theta,\bar\theta)&=&\bar\psi-\theta(i\dot{q}-b)+i\theta\bar{\theta}\dot{\bar{\psi}},\label{covbd}
\end{eqnarray}
give
\begin{eqnarray}
\int_{t_{i}}^{t_{f}}dt\int d\theta d\bar\theta D\phi(t,\theta,\bar\theta)&=&-i\int_{t_{i}}^{t_{f}}dt\dot\psi=-i\psi\big\vert_{t_i}^{t_f},\label{dertotal1}\\
\int_{t_{i}}^{t_{f}}dt\int d\theta d\bar\theta \bar D\phi(t,\theta,\bar\theta)&=&i\int_{t_{i}}^{t_{f}}dt\dot{\bar\psi}=i\bar\psi\big\vert_{t_i}^{t_f}.\label{dertotal2}
\end{eqnarray}
Hence such terms can be added to the Lagrangian without changing the equations of motion. Moreover, integration by parts can be done $\int dt d\theta d\bar\theta\,\phi_1 D\phi_2=-\int dt d\theta d\bar\theta\, D\phi_1 \phi_2+$ boundary terms, where the boundary terms, $-i(q_1\psi_2+q_2\psi_1)\vert_{t_i}^{t_f}$, can be neglected.

There are also chiral superfields which are complex
and satisfy the covariant condition $\bar{D}\phi=0$, and which can be written as $\phi(\tilde t,\theta)=A(\tilde t)+\theta\psi(\tilde t)$, where $\tilde t=t+i\theta\bar\theta$.

The superfield formalism allows to write supersymmetric actions as superspace integrals of superfield Lagrangians, taking advantage that the supersymmetry transformation of the superspace integral of a superfield is $\int d\theta d\bar\theta \delta_{\xi}\phi(t,\theta,\bar\theta)=\int d\theta d\bar\theta (\xi Q-\bar\xi\bar Q)\phi(t,\theta,\bar\theta)$, and that, similar to (\ref{dertotal1}) and (\ref{dertotal2}), 
\begin{eqnarray}
\int_{t_{i}}^{t_{f}}dt\int d\theta d\bar\theta Q\phi(t,\theta,\bar\theta)&=&i\psi\vert_{t_i}^{t_f},\label{q1}\\
\int_{t_{i}}^{t_{f}}dt\int d\theta d\bar\theta \bar Q\phi(t,\theta,\bar\theta)&=&-i\bar\psi\vert_{t_i}^{t_f}.\label{q2}
\end{eqnarray}
Therefore, a lagrangian will be a function of superfields and their covariant derivatives of first order, and the corresponding action will be of the general form
\begin{equation}
S=\int_{t_{i}}^{t_{f}}dt\int d\theta d\bar\theta L(\phi,D\phi,\bar D\phi),\label{accionsusy}
\end{equation}
whose supersymmetry transformation is
\begin{equation}
\delta_\xi S=\int_{t_{i}}^{t_{f}}dt\int d\theta d\bar\theta\, \delta_{\xi}L(\phi,D\phi,\bar D\phi)=\int_{t_{i}}^{t_{f}}dt\int d\theta d\bar\theta (\xi Q-\bar\xi\bar Q)L=\left.i\left(\xi L_\theta-i\bar\xi L_{\bar{\theta}}\right)\right\vert_{t_i}^{t_f},\label{trafos}
\end{equation}
where $L_\theta$ and $L_{\bar{\theta}}$ are the $\theta$ and $\bar{\theta}$ components of the expansion (\ref{superf}) of the superfield $L$.

The variation of the action (\ref{accionsusy}) is
\begin{eqnarray}
\delta S&=&\int_{t_{i}}^{t_{f}}dt\int d\theta d\bar\theta\left(\delta\phi\frac{\partial L}{\partial\phi}+\delta D\phi\frac{\partial L}{\partial D\phi}+\delta \bar D\phi\frac{\partial L}{\partial \bar D\phi}\right)\nonumber\\&=&\int_{t_{i}}^{t_{f}}dt\int d\theta d\bar\theta\delta\phi\left(\frac{\partial L}{\partial\phi}- D\frac{\partial L}{\partial D\phi}-\bar D\frac{\partial L}{\partial \bar D\phi}\right)+{\rm B.T.},
\end{eqnarray}
where the boundary terms are
\begin{eqnarray}
{\rm B.T.}&=&\int_{t_{i}}^{t_{f}}dt\int d\theta d\bar\theta\left[D\left(\delta\phi\frac{\partial L}{\partial D\phi}\right)+\bar D\left(\delta\phi\frac{\partial L}{\partial \bar D\phi}\right)\right]\nonumber\\
&=&-i\left.\left\{\delta q\left[\left(\frac{\partial L}{\partial D\phi}\right)_{\theta}-\left(\frac{\partial L}{\partial \bar D\phi}\right)_{\bar\theta}\right]+\delta\psi\left(\frac{\partial L}{\partial D\phi}\right)_q-\delta\bar\psi\left(\frac{\partial L}{\partial \bar D\phi}\right)_q\right\}\right\vert_{t_{i}}^{t_{f}}.
\end{eqnarray}
Therefore, it is enough if $\delta q(t_{i,f})=0$ and $\delta \psi(t_{i,f})=\delta\bar\psi(t_{i,f})=0$, which for the superfield formulation can be completed to $\delta\phi(t_i,\theta,\bar\theta)=\delta\phi(t_f,\theta,\bar\theta)=0$.
Therefore the equations of motion are
\begin{equation}
\frac{\partial L}{\partial\phi}- D\frac{\partial L}{\partial D\phi}-\bar D\frac{\partial L}{\partial \bar D\phi}=0.\label{ems}
\end{equation}
From these equations, the equations of the components are obtained from the $\theta$-expansion. These equations can be also obtained directly from (\ref{accionsusy}) written in components, i.e. after integrating the fermionic superspace variables.
\subsection{Nonconservative formulation}\label{ncsusy}
We consider supersymmetric systems under the influence of nonconservative forces \cite{previo}. We assume that these forces have a supersymmetric structure. Such systems can be described by the superfield formulation, following the lines of the bosonic formalism. As a first step, the superfields are duplicated $ \phi(t,\theta,\bar\theta)\rightarrow( \phi_1(t,\theta,\bar\theta), \phi_2(t,\theta,\bar\theta))$. Thus the noncommutative action is
\begin{equation}
S=\int_{t_{i}}^{t_{f}}dt\int d\theta d\bar\theta\left[L(\phi_1,D\phi_1,\bar D\phi_1)-L(\phi_2,D\phi_2,\bar D\phi_2)+K(\phi_1,D\phi_1,\bar D\phi_1,\phi_2,D\phi_2,\bar D\phi_2)\right],\label{accionnc}
\end{equation}
where $K(\phi_1,D\phi_1,\bar D\phi_1,\phi_2,D\phi_2,\bar D\phi_2)$ is antisymmetric under the exchange $1\leftrightarrow 2$.
The variation of this action is
\begin{eqnarray}
\delta S=\int_{t_{i}}^{t_{f}}dt\int d\theta d\bar\theta
&\Bigg\{&\delta\phi_{1}\left[\frac{\partial(L+K)}{\partial\phi_1}-D\frac{\partial(L+K)}{\partial D\phi_1}-\bar{D}\frac{\partial(L+K)}{\partial \bar{D}\phi_1}\right]\nonumber\\
&-&\delta\phi_{2}\left[\frac{\partial(L-K)}{\partial\phi_2}-D\frac{\partial(L-K)}{\partial D\phi_2}-\bar{D}\frac{\partial(L-K)}{\partial \bar{D}\phi_2}\right]\Bigg\}+{\rm B.T.}
\end{eqnarray}
The boundary terms are
\begin{eqnarray}
{\rm B.T.}=\int_{t_{i}}^{t_{f}}dt\int d\theta d\bar\theta
&\Bigg\{&D\left[\delta\phi_{1}\frac{\partial(L+K)}{\partial D\phi_1}\right]+\bar{D}\left[\delta\phi_{1}\frac{\partial(L+K)}{\partial \bar{D}\phi_1}\right]\nonumber\\
&-&D\left[\delta\phi_{2}\frac{\partial(L-K)}{\partial D\phi_2}\right]-\bar{D}\left[\delta\phi_{2}\frac{\partial(L-K)}{\partial \bar{D}\phi_2}\right]\Bigg\}.
\end{eqnarray}
Taking into account (\ref{dertotal1}) and (\ref{dertotal2}), and that $\psi=D\phi(t,0,0)$ and $\bar\psi=\bar D\phi(t,0,0)$, these boundary terms vanish if $\delta\phi_{1}(t_i,\theta,\bar\theta)=\delta\phi_{2}(t_i,\theta,\bar\theta)=0$, $\delta\phi_{1}(t_f,\theta,\bar\theta)=\delta\phi_{2}(t_f,\theta,\bar\theta)$, $\phi_1(t_f,\theta,\bar\theta)=\phi_2(t_f,\theta,\bar\theta)$, $D\phi_1(t_f,\theta,\bar\theta)=D\phi_2(t_f,\theta,\bar\theta)$ and $\bar{D}\phi_1(t_f,\theta,\bar\theta)=\bar{D}\phi_2(t_f,\theta,\bar\theta)$. Additionally, for consistency, the equality conditions of the fermionic variables at the initial time of the previous section, require the superfield conditions $\phi_1(t_i,\theta,\bar\theta)=\phi_2(t_i,\theta,\bar\theta)$. This can be seen considering that, for a supersymetric invariant theory, if one of the components of two superfields coincide, then both superfields must coincide, as can be seen from (\ref{susyt}). In components, the conditions for the superfields correspond to conditions for the components. Thus for the covariant derivatives, from (\ref{covd}) and (\ref{covbd}), we get $\dot Q_{1}(t_{f})=\dot Q_{2}(t_{f})$, $\dot\psi_{1}(t_{f})=\dot\psi_{2}(t_{f})$ and $\dot{\bar\psi}_{1}(t_{f})=\dot{\bar\psi}_{2}(t_{f})$.

Thus the equations of motion are
\begin{eqnarray}
\left(\frac{\partial}{\partial\phi_1}-D\frac{\partial}{\partial D\phi_1}-\bar{D}\frac{\partial}{\partial \bar{D}\phi_1}\right)\left[L(\phi_1)+K(\phi_1,\phi_2)\right]&=&0,\label{emsusy1}\\
\left(\frac{\partial}{\partial\phi_2}-D\frac{\partial}{\partial D\phi_2}-\bar{D}\frac{\partial}{\partial \bar{D}\phi_2}\right)\left[L(\phi_2)-K(\phi_1,\phi_2)\right]&=&0\label{emsusy2}.
\end{eqnarray}
which coincide in the physical limit, $ \phi_1(t,\theta,\bar\theta)= \phi_2(t,\theta,\bar\theta)= \phi(t,\theta,\bar\theta)$, similarly to the case of (\ref{eq1}) and (\ref{eq2}). Further, if we define the supersymmetric nonconservative forces
\begin{equation}
F_K(\phi,D\phi,\bar{D}\phi)=\left.\left(\frac{\partial}{\partial\phi_1}-D\frac{\partial}{\partial D\phi_1}-\bar{D}\frac{\partial}{\partial \bar{D}\phi_1}\right)K(\phi_1,\phi_2)\right\vert_{\phi_1=\phi_2=\phi},\label{fks}
\end{equation}
then, the nonconservative equations of motion are
\begin{equation}
\left(\frac{\partial}{\partial\phi}-D\frac{\partial}{\partial D\phi}-\bar{D}\frac{\partial}{\partial \bar{D}\phi}\right)L(\phi,D\phi,\bar{D}\phi)=-F_K(\phi,D\phi,\bar{D}\phi).\label{eqsusy}
\end{equation}
\subsection{Symmetry breaking}
Let us consider the application of Noether's theorem.
The transformations corresponding to (\ref{itema}) and (\ref{itemb}) are
$t\rightarrow t'=t+\delta t-i\xi\bar\theta-i\bar{\xi}\theta$, $\phi(t,\theta,\bar{\theta})\rightarrow\phi'(t',\theta',\bar{\theta}')+\delta_s\phi(t,\theta,\bar{\theta})+\delta_\alpha\phi(t,\theta,\bar{\theta})$, where the internal transformations $\delta_\alpha$ commute with the supersymmetry transformations, and
\begin{equation}
\delta_s\phi(t,\theta,\bar{\theta})=\delta t\dot{\phi}(t,\theta,\bar{\theta})+(\xi Q-\bar{\xi}\bar{Q})\phi(t,\theta,\bar{\theta}).\label{susyt}
\end{equation}
Therefore, for the conservative Lagrangian $L(\phi,D\phi,\bar{D}\phi)$, $\delta L=\delta_s L+\delta_\alpha L=\delta t\frac{dL}{dt}+(\xi Q-\bar{\xi}\bar{Q})L$, from which
\begin{eqnarray}
\delta L&=&
(\delta_s+\delta_\alpha)\phi\frac{\partial L}{\partial\phi}+(\delta_s+\delta_\alpha)(D\phi)\frac{\partial L}{\partial D\phi}+(\delta_s+\delta_\alpha)(\bar{D}\phi)\frac{\partial L}{\partial \bar{D}\phi},
\end{eqnarray}
which, after some arrangements, considering that the supersymmetry transformations anticommute with the covariant derivatives, and taking into account the equations of motion (\ref{ems}), turns to
\begin{eqnarray}
\delta L=D\left[(\delta_s+\delta_\alpha)\phi\frac{\partial L}{\partial D\phi}\right]+\bar{D}\left[(\delta_s+\delta_\alpha)\phi\frac{\partial L}{\partial \bar{D}\phi}\right].\label{eqnsusy}
\end{eqnarray}
Hence, taking into account the equations of motion (\ref{ems}), the conservation laws of the conservative theory of lagrangian $L$ are given by the superfield equations
\begin{eqnarray}
D\left(\dot{\phi}\frac{\partial L}{\partial D\phi}\right)+\bar{D}\left(\dot{\phi}\frac{\partial L}{\partial \bar{D}\phi}\right)-\frac{d L}{dt}&=&0,\label{nts1}\\
D\left(Q{\phi}\frac{\partial L}{\partial D\phi}\right)+\bar{D}\left(Q{\phi}\frac{\partial L}{\partial \bar{D}\phi}\right)+QL&=&0,\label{nts2}\\
D\left(\bar Q{\phi}\frac{\partial L}{\partial D\phi}\right)+\bar{D}\left(\bar Q{\phi}\frac{\partial L}{\partial \bar{D}\phi}\right)+\bar QL&=&0, \label{nts3}\\
D\left(\delta_\alpha\phi\frac{\partial L}{\partial D\phi}\right)+\bar{D}\left(\delta_\alpha\phi\frac{\partial L}{\partial \bar{D}\phi}\right)&=&0.
\end{eqnarray}
Each of these equations contains three components. Here we give the relevant components, that correspond to the energy and the supersymmetric and internal charges, which are obtained after a superspace integration
\begin{eqnarray}
E&=&-i\left(\dot{\phi}\frac{\partial L}{\partial D\phi}\right)_\theta+i\left(\dot{\phi}\frac{\partial L}{\partial \bar D\phi}\right)_{\bar\theta}-L_{\theta\bar\theta},\label{nt1}\\
J&=&-i\left(Q\phi\frac{\partial L}{\partial D\phi}-L\right)_\theta+i\left(Q\phi\frac{\partial L}{\partial\bar D\phi}\right)_{\bar\theta},\label{nt2}\\
\bar J&=&-i\left(\bar Q\phi\frac{\partial L}{\partial D\phi}\right)_\theta+i\left(\bar Q\phi\frac{\partial L}{\partial\bar D\phi}-L\right)_{\bar\theta},
\label{nt3}\\
J_\alpha&=&-iD\left(\delta_\alpha\phi\frac{\partial L}{\partial D\phi}\right)_\theta+i\bar{D}\left(\delta_\alpha\phi\frac{\partial L}{\partial \bar{D}\phi}\right)_{\bar\theta}.\label{nt4}
\end{eqnarray}
which satisfy $\dot E=0$, $\dot J=0$, $\dot{\bar{J}}=0$ and  $\dot J_\alpha=0$. 

For the nonconservative system, the transformations corresponding to (\ref{itema}) and (\ref{itemb}) are
\begin{eqnarray}
{\rm a)\ For}\ \phi_1(t,\theta,\bar{\theta}):\ \ t\rightarrow t'=t+\delta t-i\xi\bar\theta-i\bar{\xi}\theta,\ \ 
\phi_1(t,\theta,\bar{\theta})\rightarrow\phi_1'(t',\theta',\bar{\theta}')+\delta_s\phi_1(t,\theta,\bar{\theta})+\delta_\alpha\phi_1(t,\theta,\bar{\theta}).\label{trafoa}\\
{\rm b)\ For}\ \phi_2(t,\theta,\bar{\theta}):\ \ t\rightarrow t'=t-\delta t+i\xi\bar\theta+i\bar{\xi}\theta,\ \ 
 \phi_2(t,\theta,\bar{\theta})\rightarrow\phi_2'(t',\theta',\bar{\theta}')-\delta_s\phi_2(t,\theta,\bar{\theta})-\delta_\alpha\phi_2(t,\theta,\bar{\theta}),\label{trafob}
\end{eqnarray}
under which the nonconservative potential transforms as
\begin{eqnarray}
\delta K
=\delta_sK-2\left[(\delta_s+\delta_\alpha)\phi_2\frac{\partial K}{\partial\phi_2}+(\delta_s+\delta_\alpha)(D\phi_2)\frac{\partial K}{\partial D\phi_2}+(\delta_s+\delta_\alpha)(\bar{D}\phi_2)\frac{\partial K}{\partial \bar{D}\phi_2}\right].\label{deltaks}
\end{eqnarray}
Thus
\begin{equation}
\delta\Lambda=\delta_s (L_1+L_2)+\delta K,
\end{equation}
which, modulo the equations of motion (\ref{emsusy1}) and (\ref{emsusy2}), gives
\begin{eqnarray}
\delta \Lambda=D\left[(\delta_s+\delta_\alpha)\phi_1\frac{\partial \Lambda}{\partial D\phi_1}\right]+\bar{D}\left[(\delta_s+\delta_\alpha)\phi_1\frac{\partial \Lambda}{\partial \bar{D}\phi_1}\right]
-D\left[(\delta_s+\delta_\alpha)\phi_2\frac{\partial \Lambda}{\partial D\phi_2}\right]-\bar{D}\left[(\delta_s+\delta_\alpha)\phi_2\frac{\partial \Lambda}{\partial \bar{D}\phi_2}\right].\label{eqnsusy}
\end{eqnarray}
Thus, equating the right hand sides of the two preceding equations, we get
\begin{eqnarray}
&&D\left[(\delta_s+\delta_\alpha)\phi_1\frac{\partial(L_1+K)}{\partial D\phi_1}\right]
+\bar D\left[(\delta_s+\delta_\alpha)\phi_1\frac{\partial(L_1+K)}{\partial\bar D\phi_1}\right]-\delta_s(L_1+L_2+K)\nonumber\\
&&+D\left[(\delta_s+\delta_\alpha)\phi_2\frac{\partial(L_2+K)}{\partial D\phi_2}\right]
+\bar D\left[(\delta_s+\delta_\alpha)\phi_2\frac{\partial(L_2+K)}{\partial\bar D\phi_2}\right]=
-2(\delta_s+\delta_\alpha)\phi_2\left(\frac{\partial K}{\partial\phi_2}-D\frac{\partial K}{\partial D\phi_2}-\bar D\frac{\partial K}{\partial \bar D\phi_2}\right),
\end{eqnarray}
which in the physical limit gives
\begin{eqnarray}
&&D\left[(\delta_s+\delta_\alpha)\phi\frac{\partial L}{\partial D\phi}\right]
+\bar D\left[(\delta_s+\delta_\alpha)\phi\frac{\partial L}{\partial\bar D\phi}\right]-\delta_s L=
[(\delta_s+\delta_\alpha)\phi] F_K,
\end{eqnarray}
where $F_K$ is given in (\ref{fks}). Therefore, for the different transformation types we get the equations
\begin{eqnarray}
D\left(\dot\phi\, \frac{\partial L}{\partial D\phi}\right)
+\bar D\left(\dot\phi\, \frac{\partial L}{\partial\bar D\phi}\right)
-\frac{d}{dt}L&=&
\dot\phi F_K,\label{nces1}\\
-D\left(Q\phi\, \frac{\partial L}{\partial D\phi}\right)
-\bar D\left(Q\phi\, \frac{\partial L}{\partial\bar D\phi}\right)
-QL&=&Q\phi\, F_K,\label{nces2}\\
-D\left(\bar Q\phi\, \frac{\partial L}{\partial D\phi}\right)
-\bar D\left(\bar Q\phi\, \frac{\partial L}{\partial\bar D\phi}\right)
-\bar Q L&=&
\bar Q\phi\, F_K,\label{nces3}\\
D\left(\delta_\alpha\phi\, \frac{\partial L}{\partial D\phi}\right)
+\bar D\left(\delta_\alpha\phi\, \frac{\partial L}{\partial\bar D\phi}\right)&=&
\delta_\alpha\phi\, F_K.\label{nces4}
\end{eqnarray}
These superfield equations contain three components every one, which are related among them by supersymmetry transformations. The relevant components are obtained by a superspace integration, i.e. the time integral of the $\theta\bar\theta$-component, hence
\begin{equation}
\frac{dE}{dt}=\left(\dot{\phi}{F_K}\right)_{\theta\bar{\theta}}, \qquad \frac{dJ}{dt}=\left[\left(Q\phi\right) F_K\right]_{\theta\bar{\theta}},\qquad
\frac{d\bar J}{dt}=\left[ (\bar{Q}\phi) F_K\right]_{\theta\bar{\theta}}\qquad {\rm and} \qquad  
\frac{d J_\alpha}{dt}=\left[ (\delta_\alpha\phi) F_K\right]_{\theta\bar{\theta}},\label{ddtj}
\end{equation}
with $E$, $J$, $\bar J$ and $J_\alpha$ given by (\ref{nt1})-(\ref{nt4}). 
\subsection{Damped oscillator}
The conservative lagrangian and the nonconservative potential are 
\begin{eqnarray}
L(\phi,D\phi,\bar D\phi)&=&\frac{m}{2}\left(\bar D\phi D\phi+\omega_0\phi^{2}\right),\label{lcsusy}\\
K(\phi_1,D\phi_1,\bar D\phi_1,\phi_2,D\phi_2,\bar D\phi_2)&=&-i\mu/2(\bar D\phi_1 D\phi_2-\bar D\phi_2 D\phi_1).
\end{eqnarray}
The nonconservative Lagrangian is $\Lambda=L(\phi_1)-L(\phi_2)+K(\phi_1,\phi_2)$; it is a function of covariant superfields, i.e. it is supersymmetric by construction. In components the action is
\begin{eqnarray}
S=\int_{t_{i}}^{t_{f}}dt&\biggl[&\frac{m}{2}\left(\dot q_1^2-\dot q_2^2\right)-i\frac{m}{2}\left(\psi_1\dot{\bar{\psi}}_1+\bar\psi_1\dot{\psi}_1-\psi_2\dot{\bar{\psi}}_2-\bar\psi_2\dot{\psi}_2\right)+m\omega_0\left(q_1b_1+\psi_1\bar{\psi}_1-q_2b_2-\psi_2\bar{\psi}_2\right)\nonumber\\
&&+\frac{m}{2}\left(b_1^2-b_2^2\right)+\mu\left(b_1\dot{q}_2-b_2\dot{q}_1\right)+\frac{\mu}{2}\left(\bar\psi_2\dot{\psi}_1-\bar\psi_1\dot{\psi}_2+\psi_1\dot{\bar{\psi}}_2-\psi_2\dot{\bar{\psi}}_1\right)\biggl].
\end{eqnarray}
As usual, the auxiliary fields can be eliminated by their equations of motion $b_1+\omega_0 q_1+\mu/m\dot{q}_2=0$ and $b_2+\omega_0 q_2+\mu/m\dot{q}_1=0$, with the resulting action
\begin{eqnarray}
S=\int_{t_{i}}^{t_{f}}dt&\biggl[&\frac{m}{2}\left(1+\frac{\mu^2}{m^2}\right)\left(\dot q_1^2-\dot q_2^2\right)-\frac{m\omega_0^2}{2}\left(q_1^2-q_2^2\right)-i\frac{m}{2}\left(\psi_1\dot{\bar{\psi}}_1+\bar\psi_1\dot{\psi}_1-\psi_2\dot{\bar{\psi}}_2-\bar\psi_2\dot{\psi}_2\right)\nonumber\\
&&+m\omega_0\left(\psi_1\bar{\psi}_1-\psi_2\bar{\psi}_2\right)-\mu\omega_0\left(q_1\dot{q}_2-q_2\dot{q}_1\right)+\frac{\mu}{2}\left(\bar\psi_2\dot{\psi}_1-\bar\psi_1\dot{\psi}_2+\psi_1\dot{\bar{\psi}}_2-\psi_2\dot{\bar{\psi}}_1\right)\biggl].
\end{eqnarray}
The equations of motion in the physical limit are (\ref{eqfnc1}),  (\ref{eqfnc2}) and
\begin{eqnarray}
m\ddot q+ c\dot{q}+m|\zeta|^2 q=0.\label{eqbnc}
\end{eqnarray}
where $\zeta=\omega+i\mu\gamma=\frac{\omega_0}{1+\frac{\mu^{2}}{m^2}}\left(1+i\frac{\mu}{m}\right)$ (\ref{cfrec}), and $ c=\frac{2\mu}{\omega_0}|\zeta|^2$. The solutions are
\begin{eqnarray}
q(t)&=&q_0 e^{-\mu\gamma t}\cos\left(\sqrt{1-\frac{\mu^2}{m^2}}\omega t+\alpha\right),\\
\psi(t)&=&\psi_0e^{i\omega t-\mu\gamma t},\\
\bar\psi(t)&=&\bar\psi_0e^{-i\omega t-\mu\gamma t}.
\end{eqnarray}
Therefore, the frequencies of the bosonic and fermionic degrees of freedom differ. In the conservative case these frequencies coincide. This is similar to what happens in field theory after spontaneous symmetry breaking. In both cases the initial Lagrangian is supersymmetric.

The energy and the supersymmetric charges of the conservative system are
\begin{eqnarray}
E&=&\frac{m}{2}\left(1-\frac{\mu^2}{m^2}\right)\left(\dot{q}^2+\omega_0^2q^2\right)-m\omega_0\psi\bar\psi,\\
J&=&\left(i\dot q+\omega_0 q\right)\psi,\\
\bar J&=&-\left(i\dot q-\omega_0 q\right)\bar\psi,
\end{eqnarray}
where $E$, $J$ and $\bar J$ are given by (\ref{nt1})-(\ref{nt3}).
The supersymmetric nonconservative force (\ref{fks}) is $F_K=-\mu\dot\phi$. Hence from (\ref{ddtj}) we obtain the nonconservation laws
\begin{eqnarray}
\frac{dE}{dt}&=&-2\mu\left(-\omega_0\dot{q}^2+\dot{\psi}\dot{\bar{\psi}}\right),\\
\frac{dJ}{dt}&=&-\mu\left[2i\dot{q}\dot{\psi}+\omega_0\left(q\dot{\psi}-\dot{q}\psi\right)\right],\\
\frac{d\bar J}{dt}&=&-\mu\left[-2i\dot{q}\dot{\bar\psi}+\omega_0\left(q\dot{\bar\psi}-\dot{q}\bar\psi\right)\right].
\end{eqnarray}

\section{Conclusions}
We consider the application of Noether's theorem on nonconservative systems, in Galley's variational approach \cite{galley1}, to obtain the violation of the conservation laws of the corresponding conservative systems. Usually both sectors of the nonconservative action transform in the same way. We apply the inverse transformations to the time reversed sector. We assume that the nonconservative potential is symmetric under the usual transformations; in this case it is not invariant under our transformations, hence the symmetry is broken. In this framework, Noether's theorem can be applied and the right energy balance equations are obtained. 
We generalized this formalism for fermionic and supersymmetric nonconservative systems. Consistency with the first order equations of motion of fermionic variables, requires that the boundary conditions are slightly modified, with no further consequences. Otherwise the generalization is straightforward in the superfield formalism of supersymmetry. Similar to the case of time translational symmetry, for supersymmetric theories we maintain the supersymmetric structure for the nonconservative generalized potential, which is written as well in terms of superfields. The Noether theorem is evaluated and as expected, the supersymmetric charges are not conserved. We apply the results along the paper to the damped oscillator. An interesting result in the supersymmetric case, is that the frequencies of the bosonic and fermionic variables, which are the same in the conservative case, in the nonconservative case are different. This shows a parallel to the spontaneous symmetry breaking, where the initial theory is symmetric, and as a result of a locally unstable potential, the symmetry is spontaneously broken, leading to different masses for the bosonic and fermionic components.

\vskip 1truecm
\centerline{\bf Acknowledgements}
We thank VIEP-BUAP and PIFI-SEP for the support.


\end{document}